\documentclass[twocolumn,showpacs,amsmath,amssymb,aps,prb,superscriptaddress]{revtex4-1}

\pdfoutput=1
\usepackage[]{graphicx}
\usepackage[]{color}
\usepackage[dvipsnames]{xcolor}
\usepackage[breaklinks,colorlinks=true,citecolor=RoyalBlue]{hyperref}
\usepackage[]{verbatim}
\usepackage{overpic}
\usepackage{rotating}
\usepackage{mathtools}
\usepackage{appendix}
\usepackage{underscore}
\usepackage{times}
\usepackage{longtable}
\usepackage{multirow}
\usepackage{tabularx}
\usepackage[normalem]{ulem}

\makeatletter
\newsavebox{\@brx}

\newcommand{\llangle}[1][]{\savebox{\@brx}{\(\m@th{#1\langle}\)}%
  \mathopen{\copy\@brx\mkern2mu\kern-0.8\wd\@brx\usebox{\@brx}}}
\newcommand{\rrangle}[1][]{\savebox{\@brx}{\(\m@th{#1\rangle}\)}%
  \mathclose{\copy\@brx\mkern2mu\kern-0.8\wd\@brx\usebox{\@brx}}}

  \newcommand{\lllangle}[1][]{\savebox{\@brx}{\(\m@th{#1\langle}\)}%
  \mathopen{\copy\@brx\copy\@brx\mkern4mu\kern-0.7\wd\@brx\usebox{\@brx}}}
\newcommand{\rrrangle}[1][]{\savebox{\@brx}{\(\m@th{#1\rangle}\)}%
  \mathclose{\copy\@brx\copy\@brx\mkern4mu\kern-0.7\wd\@brx\usebox{\@brx}}}

\graphicspath{{./}}
\definecolor{darkblue}{RGB}{0 60 120}

\begin{document}
\newcommand\redsout{\bgroup\markoverwith{\textcolor{red}{\rule[0.5ex]{2pt}{2.0pt}}}\ULon}
\title{Realizing Haldane Model in Fe-based Honeycomb Ferromagnetic Insulators }

\author{Heung-Sik Kim}
\affiliation{Department of Physics and Center for Quantum Materials, 
University of Toronto, 60 St. George St., Toronto, Ontario, M5S 1A7, Canada}

\author{Hae-Young Kee}
\email{hykee@physics.utoronto.ca}
\affiliation{Department of Physics and Center for Quantum Materials, 
University of Toronto, 60 St. George St., Toronto, Ontario, M5S 1A7, Canada}
\affiliation{Canadian Institute for Advanced Research / Quantum Materials Program,
Toronto, Ontario MSG 1Z8, Canada}

\begin{abstract}
The topological Haldane model (THM) on a honeycomb lattice is a prototype of systems hosting
topological phases of matter without external fields. It is the simplest model
exhibiting the quantum Hall effect without Landau levels, which
motivated theoretical and experimental
explorations of topological insulators and superconductors.
Despite its simplicity, its realization in condensed matter systems has been 
elusive due to a seemingly difficult condition of  spinless fermions with 
sublattice-dependent magnetic flux terms. 
While there have been theoretical proposals including elaborate atomic-scale engineering, 
identifying candidate THM materials has not been
successful, and the first experimental realization was recently made in ultracold atoms.
Here we suggest that
a series of Fe-based honeycomb ferromagnetic insulators, 
$A$Fe$_{\bf 2}$(PO$_{\bf 4}$)$_{\bf 2}$ 
($A$=Ba,Cs,K,La) possess Chern bands described by the THM. 
How to detect the quantum anomalous Hall effect is also discussed.
\end{abstract}
\maketitle

{\color{darkblue} \it Introduction.--} 
In 1988, F.D.M. Haldane introduced an idea of the quantum Hall effect without Landau levels,
and a simple tight-binding model of spinless fermions on a 
honeycomb lattice was suggested as an example\cite{Haldane1988}, which was dubbed the { topological Haldane model} (THM).
It features a chiral edge spectrum with a Chern number without external magnetic field 
which is a prototype of the quantum anomalous Hall (QAH) insulator\cite{Weng2015,Liu2016,Ren2016}. Although THM was 
``unlikely to be directly physically realizable", as Haldane stated in his paper\cite{Haldane1988}, 
yet his vision of intrinsic topological state of matter in condensed matter systems
inspired later discoveries of time-reversal symmetric topological insulators (TI) and promoted other topological phases\cite{KaneMeleA,KaneMeleB,TI-review}. 

The THM is a spinless fermion model in a honeycomb lattice with nearest neigbor (n.n.)
and complex next nearest neighbor (n.n.n.) hopping integrals\cite{Haldane1988}:
\begin{equation}
H = t_1 \sum_{\langle ij \rangle} c^\dag_i c_j 
+ t_2 \sum_{\langle\langle ij \rangle\rangle} e^{i\Phi_{ij}} c^\dag_i c_j + h.c., \nonumber
\end{equation}
where $t_1$ and $t_2$ are real and represent n.n. and n.n.n.  hopping integral terms, respectively.
$\Phi_{ij}$ breaks time-reversal symmetry (TRS), and its sign differs for two sublattices ($A$, $B$),
i.e., $\Phi$ for $A$ and $-\Phi$ for $B$. Realization of the THM requires spinless fermions hopping 
on honeycomb lattice with spatially alternating flux yielding Aharonov-Bohm phase $i\Phi_{ij}$.

Due to these difficult requirements of the THM, 
realization of QAH effect in materials was achieved only after the discovery of
quantum spin Hall (QSH) insulator. Since each spin component of electrons in QSH insulators 
is regarded as a QAH state, one can obtain a QAH effect if one spin component QAH state is removed.
The discovery of TI\cite{TI-2008,TI-2009}, together with recent advancement of atomic-scale engineering techniques, then revived 
the interest for the QAH phase. There have been a surge of theoretical proposals in various system including magnetic-ion-doped HgTe quantum well\cite{CXLiu2008} and TI surfaces\cite{Yu2010}, 
engineered graphene\cite{ZQiao2010,Zhang2012}, transition metal oxides\cite{Xu2015,Guterding-kagome}
and their heterostructures\cite{Xiao2011,Ashley2014}. 
On the other hand, experimental observations of QAH effect was reported only in 
Cr- and V-doped (Bi,Sb)$_2$Te$_3$ film\cite{Chang2013,Chang2015},
confirming the idea that a TI with magnetic impurities removes one spin QAH state and reveals the QAH effect.
%{\color{blue}
Alternatively, the idea of breaking TRS by exerting circularly polarized ac-electric field 
and inducing QAH phase was suggested in light of the Floquet-Bloch theory\cite{Takashi2009,Kitagawa2011} 
and has been realized recently\cite{Wang453,Mahmood2016}. 
%}
Also, the experimental realization of the THM was recently made in ultracold atomic fermions 
in a periodically modulated honeycomb lattice\cite{Jotzu2014}. 
%{\color{blue}
However, it seems that realizing the THM in a simple two-dimensional honeycomb compounds
in an equilibrium situation becomes at a glance an unrealistic task. 
%}
 
Here we show that the THM, original QAH model can be found in Fe-based honeycomb ferromagnetic insulators.
With the help of strong Hund's coupling in Fe, electrons with one major spin-component (say down-spin) are fully polarized in occupied bands. Then electrons with other spin component (up-spin)
form Haldane bands with finite Chern numbers, described by
effective spinless fermions with complex n.n.n. hoppings.
We find that a series of Fe-based honeycomb stoichiometric materials, 
$A$Fe$_{\bf 2}$(PO$_{\bf 4}$)$_{\bf 2}$ ($A$FPO, $A$=Ba,K,Cs,La), fall into a class of these materials
described by the THM. Among them, compounds with $A$=K,Cs, and La
exhibit a QAH effect, while BaFe$_2$(PO$_4$)$_2$ does not show a QAH effect, because two Chern bands have opposite chirality.

\begin{figure*}
 \centering
 \includegraphics[width=0.7\textwidth]{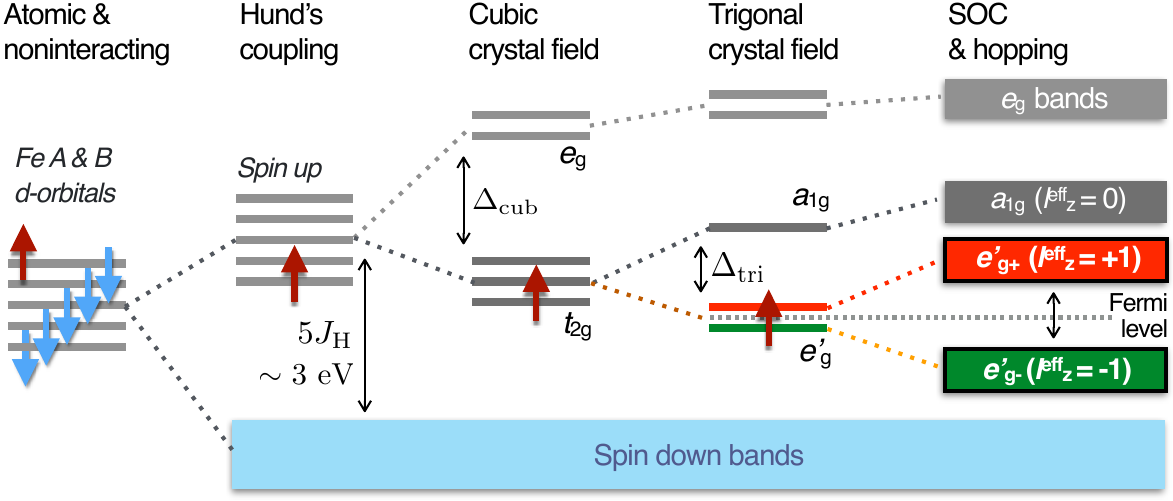}
 \caption{(Color online) Energy level diagram of Fe$^{2+}$ in BaFe$_2$(PO$_4$)$_2$ (BFPO),
 with respect to inclusion of relevant energy scales $U > J_{\rm H} > \Delta_{\rm cub}> \Delta_{\rm tri} \gtrsim  
\lambda_{\rm SO}  \approx t_{\rm hop}$.}
 \label{fig:level}
\end{figure*}

BaFe$_2$(PO$_4$)$_2$ (BFPO), a recently synthesized insulator, is the first example of two-dimensional Ising
ferromagnetic oxides, where honeycomb layers are made of FeO$_6$ octahedra
\footnote{Naming oxides is due to FeO$_6$ octahedra forming honeycomb lattice}
%AFe$_2$(PO$_4$)$_2$}.
Ferromagnetic transition occurs at $T_c \sim 65$ K, and
Fe$^{2+}$ ($d^6$) high-spin moments on honeycomb lattices align along the layer-normal direction below $T_c$.\cite{AC2012}. 
Interestingly, it also shows an intriguing re-entrant structural transition
at $T_c$ from monoclinic $P\bar{1}$ to rhombohedral $R\bar{3}$ symmetries with most likely due to
the coupling between ferromagnetic ordering and lattice structure via spin-orbit coupling (SOC).
Signature of orbital angular momentum of $\sim 1 \mu_{\rm B}$ at Fe site was reported,
implying significant role of atomic SOC in BFPO\cite{JACS2013}. It was suggested that electronic correlation 
turns the system from a semimetal to a Mott insulator with  Fe atomic orbital angular momentum\cite{Song2015}. 
We find that the atomic orbital momentum in BFPO signals possible Haldane bands via
 combined effects of Coulomb interactions and SOC under the ferromagnetic order. 
Two copies of Haldane Chern bands are identified; one set of Chern bands just above and the other set of 
Chern bands below the Fermi level, with opposite chirality of the first set. Thus BFPO is a trivial ferromagnetic Mott insulator.
However, this provides us a guideline to identify the Haldane Chern insulators, because 
Haldane Chern insulators described by the THM with effective hopping integrals can be achieved 
by adding or subtracting an electron per formula unit.
Substituting Ba into Cs, K, or La leads to new compounds 
CsFe$_2$(PO$_4$)$_2$ (CFPO), KFe$_2$(PO$_4$)$_2$ (KFPO), or LaFe$_2$(PO$_4$)$_2$ (LFPO).
%equivalent to adding one hole or electron.
We find that the structural derivatives have stable layered honeycomb structures, and exhibit
nontrivial bulk Chern numbers, featured by chiral surface states and a bulk gap of at least $\sim$ 0.2 eV, 
in the presence of Coulomb interactions of Fe $d$-orbitals.

{\color{darkblue} \it Results.--}
Fig. \ref{fig:level} shows the evolution of Fe $d$-orbital 
in BFPO with respect to inclusions of relevant energy scales. In a 3$d^6$ configuration
in the atomic limit, the dominant energy scale is the exchange splitting between 
different spin components introduced by the Hund's coupling. Density functional theory 
(DFT) calculation yields the energy difference of
$5J_{\rm H} \sim$ 3eV between the down and up spin states\cite{Song2015}, which is much larger than the
cubic crystal field of $\sim 1$ eV exerted by an oxygen octahedral cage surrounding Fe.
Hence five electrons with down-spin occupy the bands well below the Fermi level, 
and only one electron with up spin is left in the $t_{\rm 2g}$ minor spin states, which acts as
spinless fermion. 
The system without SOC becomes half-metallic
when no further on-site energy scales are included, as shown in Fig. \ref{fig:level}. 
The presence of trigonal distortion in FeO$_6$ octahedra further splits the $t_{\rm 2g}$ states into
$a_{\rm 1g}$ singlet and $e'_{\rm g}$ doublet;
\begin{align}
\vert e'_{\rm g\pm} \rangle &\equiv \frac{1}{\sqrt{3}} \left( \vert d_{xy} \rangle + e^{\pm i\theta}\vert d_{yz} \rangle + e^{\mp i\theta}\vert d_{xz} \rangle \right),  \nonumber\\
\vert a_{\rm 1g} \rangle       &\equiv \frac{1}{\sqrt{3}} \left( \vert d_{xy} \rangle + \vert d_{yz} \rangle + \vert d_{xz} \rangle \right),  \nonumber 
\end{align}
where $\theta = 2\pi/3$. Note that, the $t_{\rm 2g}$ triplet corresponds to an effective atomic orbital angular momentum $l^{\rm eff} = 1$ complex, 
$a_{\rm 1g}$ and $e'_{\rm g\pm}$ to $l^{\rm eff}_{\bf n} = 0$ and $\pm 1$ eigenstates respectively, where ${\bf n}$ is 
the layer-normal direction. Finally, in the presence of ferromagnetic order parallel to ${\bf n}$, as reported in BFPO, the spin 
degree of freedom is frozen and SOC takes the form of $\lambda_{} \hat{l}^{\rm eff}_{\bf n}\langle S_{\bf n} \rangle$, where 
the positive $\lambda_{}$ is the SOC magnitude of Fe $d$-orbital, $\hat{l}^{\rm eff}_{\bf n}$ is the effective 
atomic orbital angular momentum operator along {\bf n}, and $\langle S_{\bf n} \rangle$ is magnitude of 
the FM order. Hence SOC under the FM order behaves as an atomic orbital Zeeman field to the 
$e'_{\rm g\pm}$ states making $e'_{\rm g-}$ state lower in energy than $e'_{\rm g+}$. 
Note that, without SOC these two atomic orbital states are degenerate, and the degeneracy is
protected at $\Gamma$ and $K$ points even after hopping integrals are introduced to form the Bloch bands. 
The gap at $\Gamma$ and $K$ opens up when SOC is introduced.

The on-site Coulomb interaction enhances  the gap between the $e'_{\rm g+}$ and $e'_{\rm g-}$
Bloch states and gets bigger by the interaction parameter $U$. 
Due to the strong $d$-orbital Coulomb interaction in Fe, the system fully polarizes 
$e'_{\rm g\pm}$ orbitals,
i.e.,  the both up spin $e'_{\rm g-}$ and $e'_{\rm g+}$ 
orbitals are fully occupied and empty, respectively, as reported earlier in Ref. \onlinecite{Song2015}. 
Since the {\it effective} and {\it real} atomic orbital momenta
are antiparallel to each other, the spin and the {\it real} orbital momenta at Fe add up to yield total magnetic moment 
of $\sim 5\mu_{\rm B}$, much larger than the size of $d^6$ high-spin moment $S=2\mu_{\rm B}$  
(assuming the $g$-factor $\sim$ 2). This is consistent to the value reported in experiment\cite{JACS2013}, which
confirms that the Hubbard $U_{\rm eff}$ should be larger than 3eV, above which size of the total magnetic moment saturates close 
to the observed one. 

\begin{figure}
  \centering
  \includegraphics[width=0.4\textwidth]{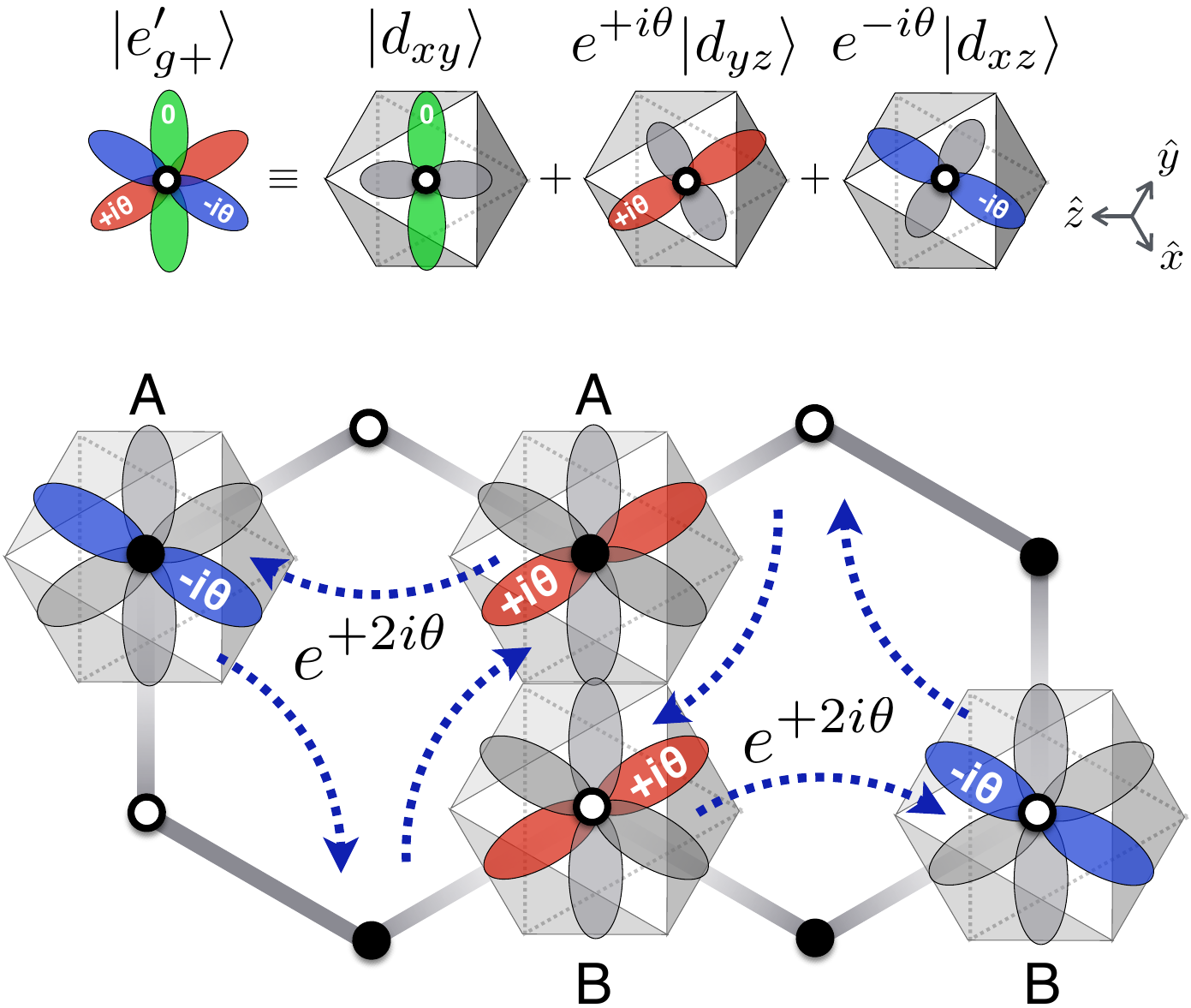}
  \caption{(Color online) Illustration of a $e'_{\rm g+}$ orbital at Fe site and
  depiction of next-nearest-neighbor (n.n.n.) complex hopping terms between the $e'_{\rm g+}$ 
  orbitals. In the hopping figure, $d_{\rm xz,yz}$ orbitals, which contribute most to the horizontal n.n.n. hopping,
  are colored within each $e'_{\rm g+}$ state. 
  Note that, the n.n.n. hopping channels represented as dashed blue arrows can be transformed to each other by the inversion
  at the n.n. bond center and the threefold rotation symmetries.  
  }
  \label{fig:hop}
\end{figure}

Let us now construct an effective tight-binding model consisting of the $e'_{\rm g+}$ orbital.
Fig. \ref{fig:hop} shows the schematic shape of $e'_{\rm g+}$ orbital. 
%{\color{blue}
First, n.n. hopping terms between the $e'_{\rm g+}$ orbitals are real-valued due to the presence of
n.n. inversion centers enforcing cancellation of complex phases. On the other hand, 
n.n.n. hopping channels can have complex values since they do not have any symmetry constraint. 
For example, let us consider one n.n.n. hopping channel between the $e'_{\rm g+}$ orbitals at A-sublattice
along the horizontal direction, as shown in lower figure in Fig. \ref{fig:hop}. 
%We assume that there is only one active n.n.n. $t_{\rm 2g}$ hopping 
%channel, connecting the $d_{yz}$ and $d_{xz}$ orbitals (highlighted in red and blue respectively 
%in the figure), and denote it as $t_{\rm off}$. 
Assuming only one off-diagonal hopping channel between the n.n.n. $t_{\rm 2g}$ orbitals is active
(between $d_{yz}$ and $d_{xz}$ orbitals, highlighted in blue and red colors in Fig. \ref{fig:hop} respectively),
in the $e'_{\rm g+}$ subspace it yields complex hopping term
$t_2 e^{+2i\theta}$, where the phase $2\theta$ originates from 
$e^{\pm i \theta}$ assigned to $d_{yz}$ and $d_{xz}$ orbitals respectively. 
%}
%The complex phases $e^{\pm i\theta}$ assigned to $d_{yz}$ and $d_{xz}$ orbitals yield complex hopping 
%components $t_2 e^{+2i\theta}$ to the n.n.n. hopping channels.
%assuming only one off-diagonal hopping channel
%between the n.n.n. $t_{\rm 2g}$ orbitals is active. 
%Note that, the complexity of the n.n.n. hopping is allowed by the absence of inversion symmetry
%at n.n.n. bond centers as shown in the figure. 
The presence of the n.n. inversion centers 
and additional three-fold symmetry then generates all other n.n.n. channels; $t_2 e^{+2i\theta}$ term for both A and B 
sublattices in a counterclockwise direction as shown in Fig. \ref{fig:hop}. For $e'_{\rm g-}$ orbitals, 
on the other hand, $t_2 e^{-2i\theta}$ terms for counterclockwise n.n.n. hopping channels are obtained. 
%{\color{blue}
The complex phase can deviate from $2\theta$ depending on details of the $t_{\rm 2g}$ 
hopping channels, but in general it does not vanish. 
%}
Hence all of the conditions for realizing THM are fulfilled in BFPO, so that the system consists of
two sets of THM ($e'_{\rm g+}$- and $e'_{\rm g-}$-THM)  with opposite chiralities to each other. 
As shown in Fig. \ref{fig:level}, the $e'_{\rm g-}$ bands are fully occupied while $e'_{\rm g+}$ bands are
empty and mixed with $a_{1g}$ band. Hence BFPO is in a trivial ferromagnetic Mott insulator phase, as
the Fermi level lies in between two sets of Haldane bands with opposite 
Chern numbers.

\begin{figure*}
  \centering
  \includegraphics[width=0.9\textwidth]{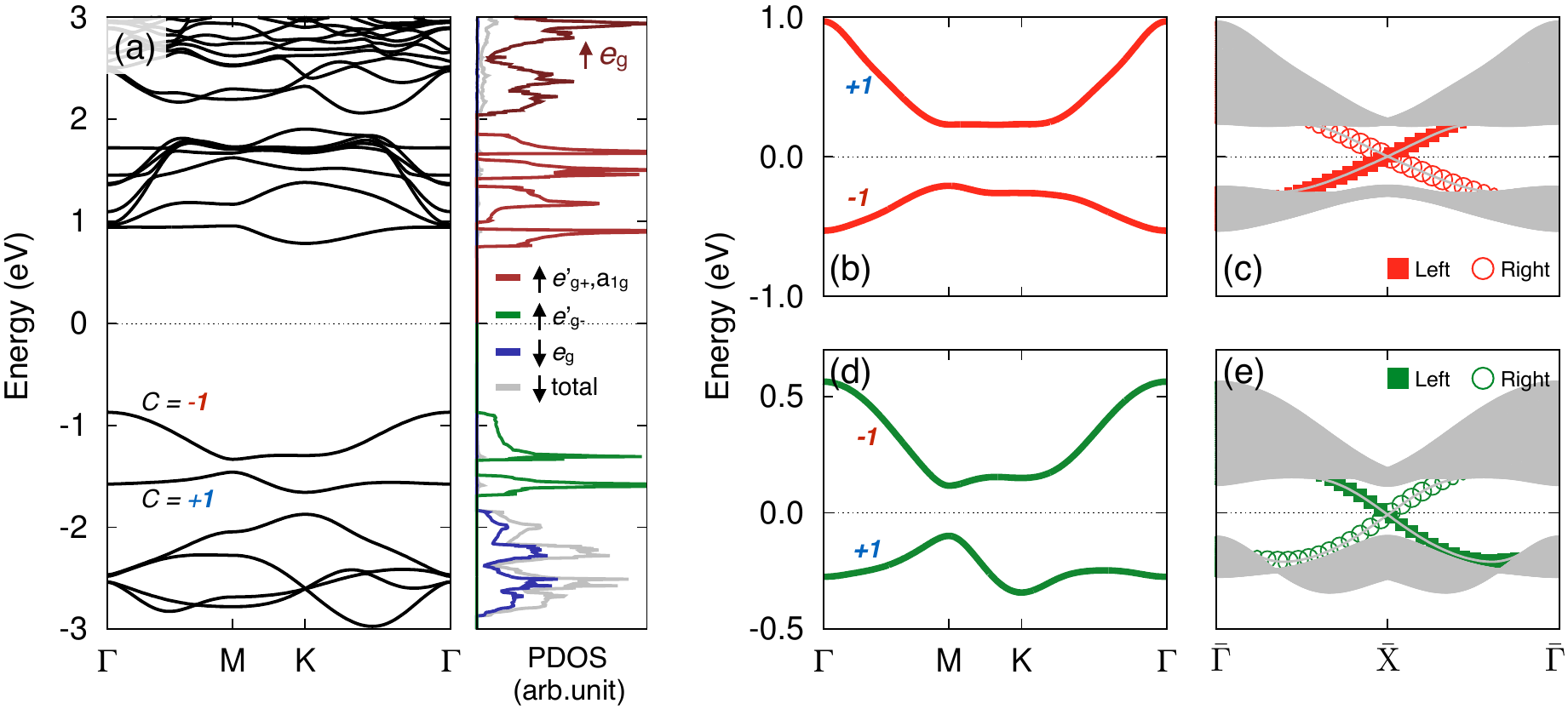}
  \caption{(Color online) (a) Band structure and projected density of states (PDOS) of 
  single-layer BFPO (SL-BFPO), where Fe up spin bands (depicted as brown, dark red, and green lines in PDOS) are mostly
  shown above -2eV with respect to the Fermi level. Chern numbers for four bands near the
  Fermi level are shown. Note that, the single-layer band structure is almost identical to the
  bulk band structure, except the position of Ba $s$-bands around 1$\sim$2 eV above the Fermi level
  as shown in Supplementary Material. $U_{\rm eff}$ = 3 eV is employed for Fe 3$d$ orbital.
  (b) Bulk $e'_{\rm g+}$ bands of SL-LFPO. 
  The occupied and unoccupied $e'_{\rm g+}$ bands are
  characterized by Chern number $C$ = -1 and +1 per a Fe$_2$(PO$_4$)$_2$ layer, respectively.
  (c) Zigzag edge spectrum of LFPO, where the size of filled square and empty circular symbols 
  depicting the edge weights. 
  (d) and (e) show the bulk and edge bands of SL-CFPO, respectively, with prevailing $e'_{\rm g-}$
  character. For LFPO and BFPO, $U_{\rm eff}$ = 4 eV is employed. 
  }
  \label{fig:U}
\end{figure*}

%{\color{blue}
The key ingredients for realizing the THM are spin-polarization due to strong Hund's coupling, 
and a separation of $e'_{\rm g+}$ and $e'_{\rm g-}$ orbitals introduced by the broken TRS
and SOC, as described above. Then, within the $e'_{\rm g+}$ (or $e'_{\rm g-}$) subspace the phase factor of 
$e^{\pm 2i \theta}$ emerges from the complex nature of the orbital wavefunction. 
To confirm our arguments above, DFT calculations incorporating Coulomb interactions and SOC are carried out.
DFT+$U$\cite{Dudarev} and Heyd-Scuseria-Ernzerhof (HSE) hybrid functional formalisms\cite{HSE06}
are employed, which yield consistent results to each other 
when the value of $U_{\rm eff}\equiv U - J$ parameter for Fe $d$-orbital
in DFT+$U$ computations is about 4 $\sim$ 5 eV
\footnote{In the weakly interacting regime ($U_{\rm eff} < 1.3$ eV) of BFPO, a narrow-gap Chern insulator/semimetal phase was 
reported previously in a DFT+$U$ study\cite{Song2016}.
The main focus of this study is how to realize the THM in the strongly interacting regime of $3 < U_{\rm eff} < 6$ eV
which is a realistic parameter range for Fe $d$-orbital. For the computational details see Supplementary Material.}. 
Fig. \ref{fig:U}(a) shows the DFT+$U$ ($U_{\rm eff}$ = 3 eV)
bands for single-layer BFPO (SL-BFPO). 
First we comment that, while the bulk unit cell contains three Fe$_2$(PO$_4$)$_2$ layers, the band splitting 
due to interlayer coupling is negligibly small as shown in Supplementary Material.  
This manifests the quasi-two-dimensional (2D) electronic 
structure of this compound, possibly one of the closest to the 2D limit among the other quasi-2D layered 
compounds ever synthesized. Hence in Fig. \ref{fig:U} SL-$A$FPO bands are shown.
The spin is oriented along the ${\bf n}$-direction, and all the down spin states 
are located below -2 eV, as shown in the projected density of states (PDOS) plot.
About 1.5 eV below and 2 eV above the Fermi level, the up spin $e_{\rm g-}$ and $e_{\rm g+}$ bands, 
respectively, are located. Chern numbers for the $e_{\rm g-}$ bands are $\pm 1$
(for bulk $\pm 3$, with each layer contributing $\pm 1$). 

While BFPO is a failed QAH though described by the THM, this system offers us an insight.
One can obtain a half-filled $e'_{\rm g\pm}$ bands by adding one electron or hole per a formula unit, and this can be achieved by
substituting Ba into alkali or lanthanide elements of similar ionic radii with Ba$^{2+}$, 
say K$^{1+}$ and La$^{3+}$ for alkali and rare-earth ions, respectively
\footnote{Applying gate voltage on two-dimensional BFPO sheets is another way to generate QAH effects and
to reveal the associated Chern bands of THM.}.
Since BFPO is a layered compound with Ba layers residing between the Fe$_2$(PO$_4$)$_2$ layers, replacing
Ba into other cations can be done by intercalation of substitute cations or by using thin-film growth technique. 

\begin{table}[t!]
  \centering
  \setlength\extrarowheight{2pt}
  \newcolumntype{R}{>{\raggedleft\arraybackslash\hsize=.6\hsize}X}
  \newcolumntype{W}{>{\raggedleft\arraybackslash\hsize=1.0\hsize}X}
   \begin{tabularx}{0.48\textwidth}{lRRRRR}
   \hline\hline   
   & $t_1$ & $t_2$ & $t'_2$ & $t_3$ & $t_{\rm inter}$  \\[5pt] \hline
  LFPO ($e'_{\rm g+}$) & -210.4   & -50.3      & +22.0     & -15.7     & -5.2  \\   
  CFPO ($e'_{\rm g-}$) & -138.5   & +46.2     & +21.4     & +5.3      & -0.2   \\
  KFPO ($e'_{\rm g-}$) & -137.6   & +46.3     & +24.0     & +3.8      & +1.2   \\
  BFPO ($e'_{\rm g-}$) & -116.7   & +24.0     & +39.1     & -7.7      &   \\
    \hline\hline 
  \end{tabularx}
  \caption{(Units in meV)
  Hopping integrals for the $e'_{\rm g\pm}$ states in bulk LFPO and CFPO/KFPO respectively,
  $e'_{\rm g\pm}$-hopping integrals for SL-BFPO is shown for comparison ($U_{\rm eff}$ = 4 eV for all systems). 
  $t_3$ and $t_{\rm inter}$ are the third-nearest-neighbor and the largest interlayer hopping terms, 
  respectively. 
}
  \label{tab:hops}
\end{table}

Next, the results for LFPO and CFPO are presented.
Structural optimizations within $R{\bar 3}$ symmetry for both CFPO, KFPO, and LFPO with including van der Waals
functionals were done to obtain the lattice parameters and internal coordinates, where the optimized structures
are shown in Supplementary Materials. 
Note that, since results for CFPO and KFPO are almost identical to 
each other, except the ${\bf c}$ parameter, here we only show results from CFPO. 
Fig. \ref{fig:U}(b) and (c) shows the LFPO bulk and zigzag-edge bands
respectively, dominated by the $e_{\rm g+}$ states ($U_{\rm eff}$ = 4 eV). The occupied and unoccupied bands have Chern number $C$ = -1
and +1 per a Fe$_2$(PO$_4$)$_2$ layer, respectively, with a well-defined bulk gap of $\sim$ 400 meV and showing a chiral edge state 
at one zigzag edge side as shown in Fig. \ref{fig:U}(c). On the other hand, Fig. \ref{fig:U}(d) and (e) shows the
CFPO bands, showing $e_{\rm g-}$ state character with opposite chirality. The bulk gap is about 220 meV,
which is smaller than that of LFPO but still substantial. The hopping integrals for the $e_{\rm g\pm}$ states
in both compounds are obtained from the Wannier orbital calculations, where the values are shown in Table \ref{tab:hops}. Note that they are well described by the THM.
Unlike in 4$d$- or 5$d$ honeycomb materials, such as $\alpha$-RuCl$_3$ or $A_2$IrO$_3$ ($A$=Li,Na) having the similar 
edge-sharing octahedral structure, $A$FPO show almost negligible third-nearest neighbor hopping terms due to the
spatially localized 3$d$ orbitals\cite{Andrei2016}. $t_3$ term in LFPO is larger than that of CFPO due to the
smaller in-plane lattice constant, and further enhancing $t_3$ with epitaxial strain may induce a phase transition from 
$C=\pm 1$ to $\pm 2$ phase as reported in $\alpha$-RuCl$_3$ or $A_2$IrO$_3$\cite{Andrei2016}.

It should be commented that,
although the size of the BFPO band gap in DFT+$U$ and HSE results are well matched at $U_{\rm eff} \simeq$ 5 eV, 
our results for all of $A$FPO systems in this work remain robust in a wide range of $U_{\rm eff}$ value, 
between 1.8 and 7 eV. 
%{\color{blue}
%In addition, the Jahn-Teller type distortion observed in a DFT calculation for BFPO\cite{JACS2013}
%is not observed in our results when the system is driven into the strongly interacting regime with SOC, 
%implying the role of orbital moment formation in maintaining the rhombohedral phase with ideal
%Fe honeycomb lattices.
%}
In CFPO (KFPO) and LFPO, a bipartite charge ordering which breaks the topological phase
is not observed across the $U_{\rm eff}$ parameter range we tested. 
%{\color{blue}
%A Fe-dimerized phase in the honeycomb plane exists as another energy minimum in KFPO, originating from the $e'_{\rm g-}$ 
%bond orbital formation, but it may survive only at very low temperature due to the low-dimensionality. 
%}
These observations suggest the robustness of THM and the resulting topological phase in these systems.

{\color{darkblue} \it Discussions.--}
Our result suggests a new strategy in realizing THM phase with sizable gap in condensed matter systems. 
Previously, there have been roughly two different approaches; depositing magnetic ions or interfacing magnetic systems
onto graphene or other honeycomb lattices\cite{Qiao2010,Qiao2014,Garrity2013}, and planting magnetic ions in topological
insulators\cite{Yu2010,Chang2013,Chang2015}.
We find that searching for systems with the ferromagnetic order in transition metal compounds with strong Hund's coupling
and SOC is a promising way for the realization of THM.
In addition to $A$FPO investigated in this study, there has been a report of possible quasi-2D Ising ferromagnets in 
several 3$d$-transition metal halides\cite{TMH1,TMH2}. Since they have a $t_{\rm 2g}$ orbital degree of freedom with spin splitting larger than
the cubic crystal field, such system
can show a similar atomic-orbital formation which may possibly lead to the formation of THM. 
Heterostructures of transition metal oxides with magnetic ions, such as a double perovskite Ba$_2$FeReO$_6$\cite{Arun2016}, 
can be another candidate. 

Lastly we comment on the size of the gap in BFPO. In the paramagnetic high-temperature $R\bar{3}$ phase, activation energy
estimation from the conductivity is about 0.2 eV\cite{David2014}, while optical spectroscopy estimates the 
gap size to be about 1.5 eV\cite{David2014}, which are significantly smaller than our results of $\sim$3 eV
from the hybrid functional calculation. More reliable low-temperature data should be measured for further
studies, and we emphasize that our results is robust independent of the value of $U$ and corresponding gap,
as long as $U_{\rm eff}$ is larger than 1.8eV.
A further theoretical studies incorporating quantum fluctuation effects for these systems when a fraction of electron or hole
is added would be interesting, which may reveal fractional Chern insulator phases in these systems\cite{Neupert2011}. 

In summary, we propose a way to search for realistic materials described by the THM:
effectively spinless fermion  with
complex n.n.n hopping integrals can be found in ferromagnetic insulators with strong Hund's coupling
and finite SOC. 
We apply this idea to a series of Fe-based honeycomb ferromagnetic oxides. We show that
$A$FPO series is represented by THM, and predict that honeycomb CFPO, KFPO, and LFPO are candidates for the original THM 
exhibiting a quantum Hall effect without Landau levels. 
Our study provides a platform for exploring correlated topological materials, 
including possible fractional Chern insulators via strong correlations.

{\color{darkblue} \it Methods.--}  DFT computations in this work are done with 
Vienna {\it ab-initio} Simulation Package ({\sc vasp})\cite{VASP1,VASP2} and {\sc openmx} 
codes\cite{openmx}. Especially structural optimization for C/KFPO and LFPOs with van der Waals functionals,
and HSE hybrid functional calculations are done with using {\sc vasp}, and in DFT+$U$ calculations 
and Wannier orbital calculations {\sc openmx} was employed. DFT+$U$ results from {\sc openmx}
and {\sc vasp} are compared and found to be consistent with each other. For more computational details
see Supplementary Material. 

{\it Acknowledgements:}
HSK thanks Ji-Sang Park, Andrei Catuneanu, and Yige Chen for helpful comments. 
This work was supported by the NSERC of
Canada and the center for Quantum Materials at the University of
Toronto.  Computations were mainly performed on the GPC supercomputer
at the SciNet HPC Consortium. SciNet is funded by: the Canada
Foundation for Innovation under the auspices of Compute Canada; the
Government of Ontario; Ontario Research Fund - Research Excellence;
and the University of Toronto.

\bibliography{BFPO}

\clearpage

\section{Supplementary Material A:\\
Computational details}

For the structural optimization, we employ 
the Vienna {\it ab-initio} Simulation Package ({\sc vasp}), which uses the 
projector-augmented wave (PAW) basis set\cite{VASP1,VASP2}. 
520 eV of plane wave energy cutoff is used, and for
$k$-point sampling 15$\times$15$\times$15 grid including Gamma point
is employed for the rhombohedral primitive cell. A revised Perdew-Burke-Ernzerhof (PBE) generalized gradient 
approximation (PBEsol)\cite{PBEsol} is used for structural optimization and total energy 
calculations, which yields the best agreement of calculated lattice parameters compared to
local density approximation or other GGA functionals. 

Structural optimizations using {\sc vasp} are carried out in two stages. First, optimizations of 
cell parameters and internal coordinates are performed under the $R\bar{3}$ symmetry constraint and
in the absence of SOC. After that, the layer-normal ${\bf c}$-lattice parameter in Fig. \ref{figA:str} is again optimized 
with using vdW-optB86b van der Waals functional\cite{vdW-optB86b} and with fixed in-plane 
${\bf a}$-lattice parameter. vdW-optB86b results yield 2\% smaller and 0.5\% larger value of 
${\bf c}$-parameter for KFPO and LFPO, respectively, compared to PBEsol-optimized ${\bf c}$ values. 
%Lastly, internal coordinates are optimized 
%with turning on SOC and employing Dudarev's rotationally invariant DFT+$U$ formalism\cite{Dudarev}
%($U_{\rm eff}$ = 5 eV) but without any symmetry constraints. 
%The value of $U_{\rm eff}$ is chosen 
%because it gives a good agreement of the gap size between the DFT+$U$ and HSE results in BFPO, 
%as mentioned in the main text. 
Force criterion of 1 meV / \AA~is
used for internal coordinates and stress tensor optimizations. 
%Optimized structural parameters
%are refined using {\sc findsym} package\cite{Findsym} with tolerance of $5\times 10^{-4}$\AA,
%as shown in Table \ref{tabA:coord}. 

Electronic structure calculations with including Coulomb interactions are both performed 
with using {\sc vasp} and {\sc openmx}\cite{openmx} codes. In both codes, Dudarev's rotationally invariant 
DFT+$U$ formalism\cite{Dudarev} is employed with effective $U_{\rm eff} \equiv U-J$ increased up to 6 eV
for Fe $d$ orbitals and 8 eV for La $f$-orbitals. 
For hybrid functional calculations, the exchange-correlation functional of Heyd-Scuseria-Ernzerhof\cite{HSE06}
implemented in {\sc vasp} is employed, with the mixing parameter $\alpha$ = 0.25 and the inverse effective 
screening length $\omega = 0.2$ \AA$^{-1}$. For the HSE calculations, to reduce the computational cost, 
two-dimensional unit cells with vacuum of 20 \AA~and 11$\times$11$\times$1 of $k$-grid sampling are 
employed for each system. Maximally localized Wannier orbital formalism\cite{Wannier1,Wannier2}
as implemented in {\sc openmx}\cite{Wannier3} is employed for computations of 
the $e'_{\rm g\pm}$ Wannier orbitals.

\begin{figure}[ht!]
  \centering
  \includegraphics[width=0.45 \textwidth]{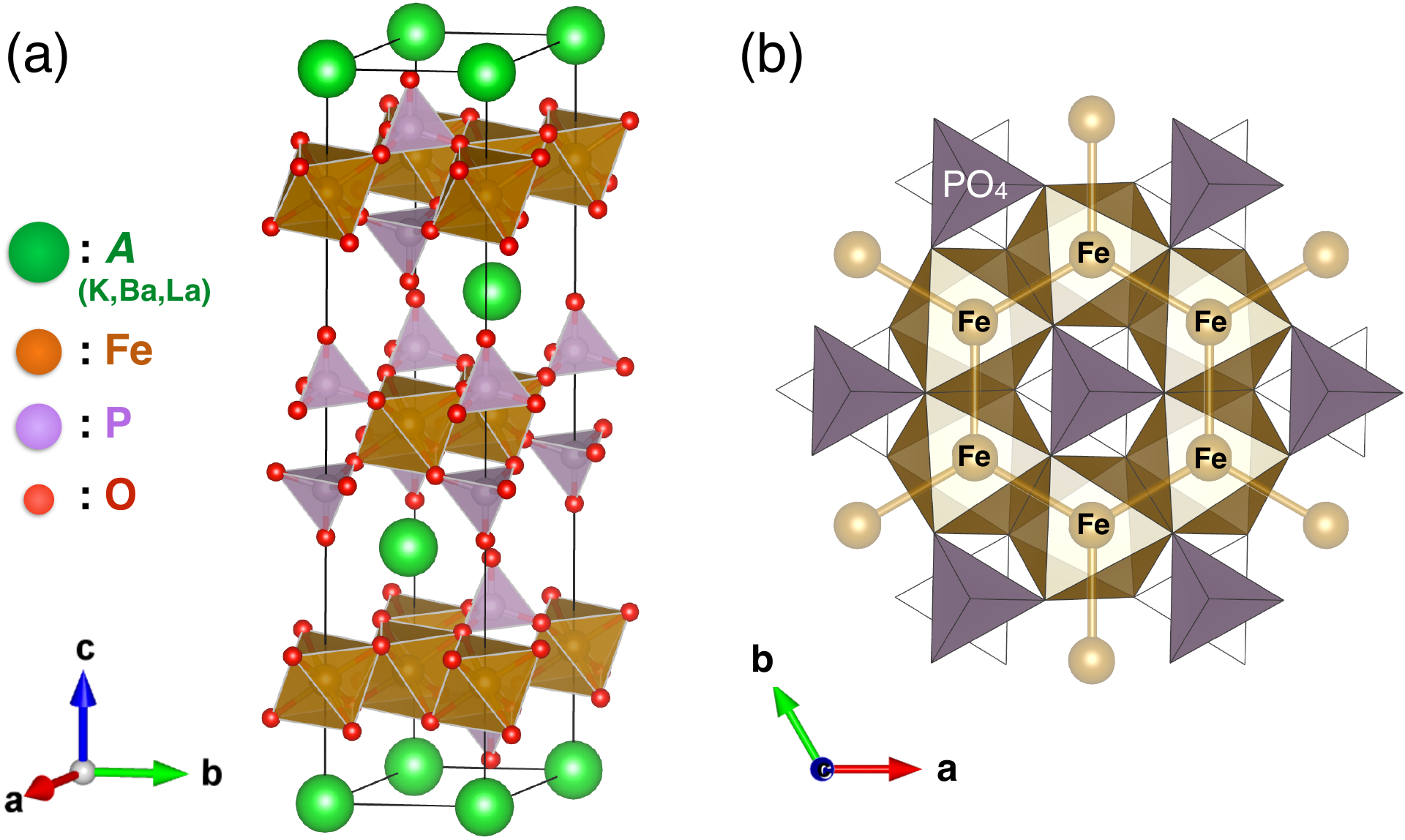}
  \caption{(Color online) 
  (a) Conventional unit cell of $A$FPO structure, and (b) the top view of a Fe$_2$(PO$_4$)$_2$ layer.
  }
  \label{figA:str}
\end{figure}

\renewcommand*{\arraystretch}{1.3}
\begin{table}
\centering
\begin{tabular}{llrrrrrr}  \hline\hline
&                        & BFPO & BFPO & BFPO & KFPO  &CFPO & LFPO\\
&                        & (T = 293K) & (1.8K) & (opt.) & (opt.) & (opt.) & (opt.)\\ \hline
& {\bf a}           & 4.873    & 4.869    & 4.824    & 4.800    & 4.831    & 4.732 \\
& {\bf c} (\AA) & 23.368 & 23.230   & 23.542 &  25.231 & 27.612  & 21.986  \\[5pt]
Fe   ($6c$)  & $z$ & 0.1701 & & 0.1698 & 0.1676 &0.1672& 0.1707 \\[5pt]
P     ($6c$)  & $z$ & 0.4243 & & 0.4244 & 0.4311 &0.4364& 0.4167 \\[5pt]
O1 ($6c$)  & $z$ &  0.6416  & & 0.6403 & 0.6268 &0.6176& 0.6538 \\[5pt]
O2 ($18f$)  & $x$ & 0.0258  && 0.0220 & 0.0269 &0.0289& 0.0139 \\
                      & $y$ & 0.3716 && 0.3675 & 0.3634 &0.3637& 0.3667 \\
                      & $z$ & 0.2208 & & 0.2204 & 0.2123 &0.2089& 0.2289 \\ \hline\hline
\end{tabular}
\caption{Table of experimental and optimized lattice parameters of $A$FPO series with $R\bar{3}$ symmetry, where
experimental lattice parameters are from Ref. \onlinecite{AC2012} and \onlinecite{JACS2013}. 
$A$-site cations (Ba,K,Cs,La) are located at $3a$ Wyckoff position (0,0,0) for all compounds, while Fe, P, and O1 at
(0,0,$z$) and O2 at ($x$,$y$,$z$). Experimental internal coordinates of BFPO at T = 1.8 K are unavailable.
}
\label{tabA:coord}
\end{table}

\section{Supplementary Material B:\\
optimized coordinates for $A$FPO}

Table \ref{tabA:coord} presents the optimized structure for $A$FPO systems in the $R\bar{3}$ symmetry, compared to
the experimentally reported structures. Cell parameter optimizations for BFPO yields
0.9\% smaller and 1.3\% larger ${\bf a}$ and ${\bf c}$ parameters, respectively,
compared to the experimentally observed ones at T = 1.8K. Optimized BFPO internal coordinates 
%in the DFT+$U_{\rm eff}$ calculation 
show fairly good agreement to the observed one at T = 293K.
%compared to the PBEsol-only optimized ones,
%especially in the increased size of Fe honeycomb lattice buckling
%from the ideal 2D one, {\it i.e.} deviation of the Fe position from $z$ = 1/6 in 
%Table \ref{tabA:coord}. 
FeO$_6$ octahedral distortion is slightly larger compared to the room-temperature structure. 
In KFPO and LFPO, the ${\bf c}$ parameter is increased and decreased by 7\% compared to BFPO,
respectively, reflecting the ionic character of the interlayer bonding mediated by the 
K$^{1+}$, Ba$^{2+}$, or La$^{3+}$ cations. Compared to KFPO, CFPO shows 10\% larger
${\bf c}$ parameter due to the larger ionic radius of Cs$^{1+}$ compared to K$^{1+}$,
but without significant differences in the ${\bf a}$ parameter size and internal coordinates. 
KFPO and CFPO shows smaller octahedral distortion and Fe buckling than BFPO, with small mismatch of 
the in-plane lattice constant compared to the BFPO in-plane lattice constant
(0.5\% and 0.1\% for KFPO and CFPO, respectively). 
LFPO, on the other hand, show 2\% smaller in-plane 
lattice parameter, which is attributed to the completely filled three $t_{\rm 2g}$ bonding orbitals
as shown later in Fig. \ref{figA:bands}(g). The smaller value of ${\bf c}$ parameter in LFPO yields 
increased interlayer hopping terms compared to BFPO, CFPO, and KFPO, but not enough to break the 
quasi-two-dimensional picture. 

%{\color{blue}
%It should be commented that, in the structural optimizations for KFPO and CFPO with SOC+$U_{\rm eff}$
%(without symmetry constraints), another stable phase with Fe dimerization in the unit cell is observed
%with the total energy lower by $\sim 0.2$ eV per a formula unit than the $R\bar{3}$ symmetric one. 
%This can be attributed to the formation of the bond-orbital order between the Fe $e'_{\rm g-}$ orbitals
%when there are no thermal or quantum fluctuations. Due to the low-dimensionality in this system, such bond orbital
%order may only survive at very low temperature due to thermal fluctuation. On the other hand,
%in LFPO, this dimerized phase is not observed due to the smaller in-plane lattice constant and the lower
%energy gain. 
%}

\begin{figure*}
  \centering
  \includegraphics[width=1.0 \textwidth]{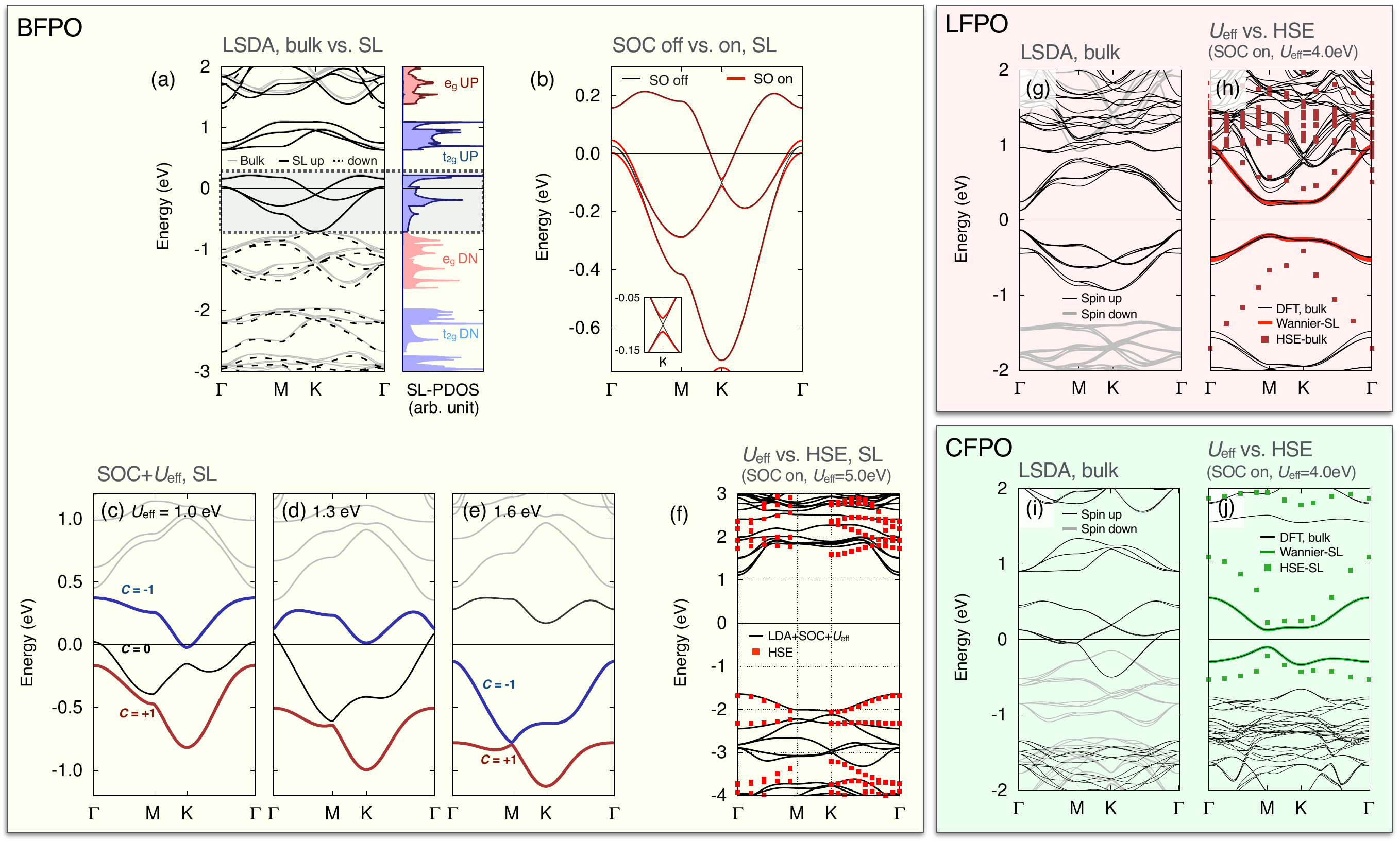}
  \caption{(Color online) 
  Band structures of BFPO (a-f), LFPO (g,h), and CFPO (i,j), comparing its evolution 
  with respect to inclusion of SOC and Fe $d$-orbital Coulomb interactions. 
  (a) shows the spin-polarized band structure and projected density of states (PDOS) of BFPO 
  without SOC, comparing the bulk and single-layer (SL) results. The black solid and dashed lines
  depict the spin up and down SL bands, respectively, while grey lines representing bulk ones. 
 PDOS is from the SL result. (b) Comparison of the bands without and with SOC. The SOC-induced 
 gap size at K point is about 30 meV, as shown in the inset. (c-e) Evolution of the bands 
 as a function of increasing $U_{\rm eff}$ value. $U_{\rm eff}$ = 1.3 eV is the critical value
 of the Chern-to-Mott insulator transition in our work. 
  (f) Comparison between the DFT+$U_{\rm eff}$ ($U_{\rm eff}$ = 5 eV)  and the HSE bands,
   where the HSE bands are depicted as the red square symbols. 
   (g) shows the LFPO bulk spin-polarized bands without SOC, and (h) shows the bands with including
   SOC and $U_{\rm eff}$ = 4 eV. HSE bands depicted as red square symbols are overlaid. 
  (i) and (j) shows the bulk CFPO bands without and with the Coulomb interactions, respectively
  (green squares depicting HSE results). 
  }
  \label{figA:bands}
\end{figure*}

\section{Supplementary Material C:
Electronic structure results}

Band structure and its evolution with respect to the inclusion of $U$ in BFPO is presented in 
Ref. \onlinecite{Song2015} and \onlinecite{Song2016}. Our result, summarized in Fig. \ref{figA:bands}(a-f)
is consistent with theirs. Here we comment some points which weren't mentioned in previous studies. First, the coupling between
the different Fe$_2$(PO$_4$)$_2$ is very weak in BFPO. Fig. \ref{figA:bands}(a) shows the bulk BFPO bands
in a hexagonal unit cell with three Fe$_2$(PO$_4$)$_2$ layers (grey lines) overlaid with a single-layer (SL)
BFPO bands (black solid and dashed lines). Band splitting due to interlayer coupling is almost absent in the
bulk bands, smaller then the linewidth to draw the SL bands in the plot. As shown later, this quasi-2D character
is maintained in CFPO, KFPO, and BFPO, as mentioned in the previous section. 
Next, the three bands near the Fermi level in the absence of SOC and $U$, enclosed in a shaded box 
in Fig. \ref{figA:bands}(a), consist of $t_{\rm 2g}$ bonding-orbitals centered at each n.n. bond center. 
This bonding-orbital formation is driven by a strong $\sigma$-like $dd$ direct overlap between the
two n.n. $t_{\rm 2g}$ orbitals, where the overlap integral is about -0.45 eV and overcomes other 
$t_{\rm 2g}$ hopping channels. The prevailing $\sigma$-overlap occurs due to the edge-sharing geometry 
of FeO$_6$ octahedra yielding rather short n.n. distance. 
Hence BFPO in the weakly correlated limit forms a bonding-orbital kagome lattice, as suggested 
in Ref. \onlinecite{sd2-graphene}, but with significant n.n.n. and n.n.n.n. hopping terms. 
Degeneracies at $\Gamma$ and $K$ points are protected by the complex conjugation 
$\mathcal{K}$ and the product of inversion and $\mathcal{K}$ respectively, where $\mathcal{K}$ is a 
TRS operation followed by a global spin rotation reorienting the spin moments to the original direction. 
SOC breaks this pseudo-TRS and opens the gap at both points, as shown in Fig. \ref{figA:bands}(b). 

As mentioned in the main text, the local atomic orbital picture survives in the strongly correlated regime. 
Hence a crossover happens in a intermediate strength of $U$ (or equivalently $U_{\rm eff}$). 
Fig. \ref{figA:bands}(c-e) shows the crossover, where the transition from a Chern to a trivial Mott insulator 
happens at $U^c_{\rm eff}$ = 1.3 eV in our calculations\cite{Song2016}. This critical value is smaller than 
the one reported in Ref. \onlinecite{Song2015}, $U^c - J$ = 2.5 - 0.7 = 1.8 eV, due to the difference in the choice of 
correlated orbital projectors. The loss of the Chern insulator phase and the onset of the large orbital moment
formation is a signature of the $e'_{\rm g\pm}$-polarized atomic orbital picture. Just above the $U^c_{\rm eff}$, the 
lowest unoccupied band carries a zero Chern number, however it crosses again with other $t_{\rm 2g}$ bands
and obtains a nontrivial Chern number in higher $U_{\rm eff}$ value. 
%{\color{blue}
Note that, in Ref. \onlinecite{Song2016}, the bulk Chern number in the weak-$U$ regime was -3. 
While the sign difference originates from the direction of the magnetic moments, the difference in 
the Chern number magnitude merely comes from different definition. In Ref. \onlinecite{Song2016}, 
the total Chern number (per unit cell) is used, while in our paper it is the Chern number per a FePO$_4$ layer.
The Chern number magnitude in Ref. \onlinecite{Song2016} is consistent with ours ($\pm$1 per a FePO$_4$ layer) 
since the bulk primitive
unit cell with rhombohedral symmetry employed in Ref. \onlinecite{Song2016} implies the presence of
three FePO$_4$ layers in the unit cell. Alternatively a conventional hexagonal unit cell including three
FePO$_4$ layers can be used, and both choices of bulk unit cell yield $\pm$3 
(sign depending on the direction of the magnetic moment) of total Chern number in the weak-$U_{\rm eff}$ regime. 
%}

Fig. \ref{figA:bands}(f) compares the DFT+$U_{\rm eff}$ result ($U_{\rm eff}$ = 5 eV) to the HSE result 
in SL-BFPO. The DFT+$U_{\rm eff}$ band gap agrees best with the HSE one in the range of $U_{\rm eff}$ = 4.5 
to 5 eV, and we chose 5 eV for presentation of the results. HSE calculation reproduces the formation of 
$e'_{\rm g\pm}$-polarized band structure, as shown in Fig. \ref{figA:bands}(f). Compared to DFT+$U_{\rm eff}$,
HSE bands show larger exchange splitting between the $e'_{\rm g-}$ and the fully occupied spin states,
yielding well-separated $e'_{\rm g-}$ THM bands from the others. Also, the $s$-like bands at the bottom of 
conduction bands are pushed above in the HSE result. 

Now the bulk LFPO (Fig. \ref{figA:bands}(g,h)) and CFPO (Fig. \ref{figA:bands}(i,j)) results are presented. 
Note that, KFPO bands are almost identical to those from CFPO and are not presented here. 
In LFPO, in the absence of SOC and $U_{\rm eff}$, the $t_{\rm 2g}$ bonding orbital bands are fully
occupied, as shown in Fig. \ref{figA:bands}(g). This induces the smaller in-plane lattice constant in LFPO 
compared to the others. Note that, the band splitting due to the interlayer coupling is visible, 
although not significant, due to the smaller interlayer distance here. Inclusion of $U_{\rm eff}$ 
= 4 eV transforms the orbital landscape from the bonding-antibonding picture to the atomic one,
and in Fig. \ref{figA:bands}(h) the half-filled $e'_{\rm g+}$ THM bands are clearly seen. 
Interestingly, HSE result in Fig. \ref{figA:bands}(h) shows larger $e'_{\rm g+}$ THM bandwidth 
with larger band gap of $\sim 0.8$ eV at the K point, with much better separated $e'_{\rm g+}$ THM bands 
compared to the DFT+$U_{\rm eff}$ result. Similarly, CFPO HSE result in Fig. \ref{figA:bands}(j) shows
much larger separation between the $e'_{\rm g+}$ THM bands the lower ones, and the larger bandwidth
and gap size of the THM bands. These enhancement of gap sizes and bandwidths, compared to the
DFT+$U_{\rm eff}$ results, might be attributed to the nonlocal correlation effect inherent in the
hybrid functional approach but absent in DFT+$U_{\rm eff}$.

\end{document}